\documentclass{article}

\usepackage{arxiv}

\usepackage[utf8]{inputenc} 
\usepackage[T1]{fontenc}    
\usepackage{hyperref}       
\usepackage{url}            
\usepackage{booktabs}       
\usepackage{amsfonts}       
\usepackage{nicefrac}       
\usepackage{microtype}      
\usepackage{lipsum}		
\usepackage{graphicx}
\usepackage{natbib}
\usepackage{doi}

\title{Neural-ISAM: A hybrid in-situ machine learning approach for complex manifold-based combustion models in LES of turbulent flames}

\author{\hspace{1mm}S.~Trevor Fush\thanks{Corresponding author, email: sf5201@princeton.edu} \\
	Department of Mechanical and Aerospace Engineering\\
	Princeton University\\
	Princeton, NJ \\
	\And
	Israel J.~Bonilla \\
	Department of Mechanical and Aerospace Engineering\\
	Princeton University\\
	Princeton, NJ \\
	\AND
    Michael B.~Schroeder \\
	Department of Mechanical and Aerospace Engineering\\
	Princeton University\\
	Princeton, NJ \\
    \AND
    Matthew X.~Yao \\
    Department of Mechanical Engineering\\
    University of New Brunswick\\
    Fredericton, Canada\\
	Department of Mechanical and Aerospace Engineering\\
	Princeton University\\
	Princeton, NJ \\
    \AND
    Michael E.~Mueller \\
	Department of Mechanical and Aerospace Engineering\\
	Princeton University\\
	Princeton, NJ \\
}

\renewcommand{\shorttitle}{Neural-ISAM}

\begin{document}
\maketitle

\begin{abstract}
	Manifold-based combustion models are used to decrease the cost of turbulent combustion simulations by projecting the thermochemical state onto a lower-dimensional manifold, allowing the thermochemical state to be computed separately from the flow solver. The solutions to the manifold equations have traditionally been precomputed and pretabulated for use in the simulations, but this results in large memory requirements and significant precomputation cost even for simple models. One approach to alleviate the memory requirements is to use In-Situ Adaptive Manifolds (ISAM), which only stores manifold solutions that are encountered during a simulation in a database built with In-Situ Adaptive Tabulation (ISAT). Even with ISAM, as the manifold complexity increases, the memory requirements can still grow too large. Another promising approach to reduce memory requirements of manifold databases are machine learning methods (typically neural networks), for they can represent complicated functions in a highly memory-compact manner. However, the current implementation of machine learning methods still require the generation of a training dataset without any knowledge of the conditions that will be encountered in a simulation. This work develops the Neural In-Situ Adaptive Manifolds (Neural-ISAM) method, which builds off of the original ISAM method. Neural-ISAM is designed to address the drawbacks of both adaptive tabulation and machine learning methods, while also leveraging their benefits by coupling neural networks to the manifold databases on-the-fly. ISAM databases are built via ISAT, which stores the manifold solutions in a binary tree, and Neural-ISAM periodically searches this binary tree to identify regions that can be pruned. Neural networks are trained on the candidate prune regions, and these portions of the binary tree are then replaced by the trained neural network, reducing the memory requirements of the database. The Neural-ISAM method is evaluated with LES in two turbulent flames with increasing manifold model complexity: Sandia Flame D and the Sandia Sooting flame. Memory usage, computational performance, and accuracy of the Neural-ISAM method are considered for each case.
\end{abstract}


\section{Introduction}
The simulation of realistic combustion systems using fully resolved Direct Numerical Simulation (DNS) is challenging for many reasons, most notably due to the large range of length and time scales that need to be resolved. Large Eddy Simulation (LES) can be leveraged to reduce the computational cost by resolving large scale and modeling small scale flow features, but the combustion physics still require the transport of a large number of species and the calculation of a large number of chemical source terms for realistic fuels \cite{lu2009toward}. To decrease the computational cost of the combustion physics, manifold-based combustion models decouple the combustion from the flow solver by projecting the thermochemical state from physical coordinates to manifold coordinates \cite{peters1984laminar,van2002flamelet,pope2013small,mueller2020physically}, which significantly decreases the number of transported quantities in the flow solver. The implementation of manifold-based models in LES traditionally requires the manifold solutions to be computed and tabulated a priori such that any state encountered during the simulation could be interpolated from the table. However, as the complexity of the manifold increases, the memory requirements for these precomputed tables increases substantially, quickly becoming impractical. Many techniques have been developed to accommodate such tables in LES, such as hybrid MPI/OpenMP parallelization \cite{mueller2012model} or MPI+MPI programming \cite{weise2015reducing}, but these methods still require the precomputation and storage of states that might not be encountered in the simulation. Additionally, these approaches still have limits for the complexity of the manifold models that can be used because the memory usage of the tables are not being reduced, instead just being distributed over more memory on the computational hardware.

To actually reduce the memory requirements of the precomputed tables, one approach that is increasing in popularity is replacing the precomputed tables with machine learning models (often neural networks) that can represent the table in a compact form. Machine learning has been widely used for reducing tabulation memory for numerous manifold(-like) combustion models, including the Flamelet Progress Variable (FPV) approach \cite{ihme2009optimal,hansinger2022deep}, Flamelet Generated Manifolds (FGM) \cite{zhang2020artificial,zhang2020large,li2023combining,salunkhe2023physics}, and the nonpremixed flamelet model \cite{emami2012laminar}. Approaches considered have included traditional neural networks \cite{ihme2008generation,ihme2009optimal,zhang2020artificial,zhang2020large,emami2012laminar,li2023combining}, Residual Neural Networks (RNNs)~\cite{hansinger2022deep}, Random Forests (RFs)~\cite{li2023combining}, Gradient Boosted Trees (GBTs)~\cite{li2023combining}, and physics informed learning~\cite{salunkhe2022chemtab,salunkhe2023physics}. In general, these works all demonstrate the significant reduction in memory requirements necessary for storing a machine learning model in place of a traditional table. However, the weakness is that the machine learning models must be trained on a pregenerated table, requiring the storage of either the table or all of the pregenerated data to train the model in the first place. Precomputation of all possible states required to train the model may not always be feasible for high-dimensional models.

To avoid the precomputation of states that are not necessarily encountered in the simulation for further memory reduction, In-Situ Adaptive Manifolds (ISAM) was developed and has shown a reduction the memory requirements of manifold databases while maintaining steady-state cost of LES \cite{lacey2021situ} even for complex, high-dimensional for multi-modal combustion \cite{novoselov2021large} or non-adiabatic combustion \cite{yao2025situ}. With ISAM, manifold solutions are computed on-the-fly when needed and then stored for reuse via In-Situ Adaptive Tabulation (ISAT) \cite{pope1997computationally, lu2009improved}. However, even with the reduced tabulated space, the memory requirements of these tabulation methods can also prove to be prohibitive for increasing manifold dimension and the inclusion of more complex combustion physics.

To address these trade-offs, this work introduces a novel approach, Neural-ISAM, which leverages the strengths of both adaptive tabulation and machine learning methods while addressing each of their weaknesses. Neural-ISAM initially operates the same as the original ISAM adaptive tabulation approach, but, as the memory of the manifold database grows, neural networks are trained in-situ on portions of the database, and these regions of the database are replaced with neural networks to reduce its memory usage. This addresses the weaknesses of previous approaches as it allows the neural network to be trained on the data that is generated during the simulation (not pregenerated data), while the manifold database memory can be reduced using the compact representation of the neural network. In addition, Neural-ISAM allows the neural networks to only learn a small portion of the input parameter space as opposed to the entire parameter space, reducing the network size and training time needed.

In this work, Neural-ISAM is presented in detail, and its performance is evaluated with LES in two canonical turbulent flames with varying manifold model complexity: Sandia Flame D \cite{barlow1998effects} and the Sandia Sooting flame \cite{zhang2011design}. The memory usage, computational cost, and accuracy of the Neural-ISAM framework are evaluated against the performance of the traditional ISAM framework, which has been previously implemented and validated for each of these flames \cite{lacey2021situ, yao2025situ}.

\section{Hybrid In-Situ Machine Learning for Manifold Models}

Both ISAM and Neural-ISAM rely on ISAT~\cite{pope1997computationally}, which is discussed first. A brief overview of ISAM~\cite{lacey2021situ} is then provided followed by the new Neural-ISAM approach. Further details of the specific manifold models used in this work are provided in Sec. \ref{ssec:test}.

\subsection{In-Situ Adaptive Tabulation (ISAT)} \label{ssec:isat}

The core data structure of an ISAT database is a binary tree that is constructed as the database is queried. To illustrate typical ISAT operation, consider a generic function $\mathbf{f}(\mathbf{x})$, where both $\mathbf{f}$ and $\mathbf{x}$ can be vectors of any dimension. Each leaf of the binary tree is associated with a specific value of the input vector $\mathbf{x}^0$, stores a solution to $\mathbf{f}(\mathbf{x}^0)$ and its Jacobian matrix $\mathbf{J}^0$, and possesses an Ellipsoid of Accuracy (EOA). The EOA represents a region within which a linear extrapolation of the solution can be performed to within a user-specified error tolerance. When the database is queried with a point $\mathbf{x}^q$, the binary tree is traversed to locate a nearby leaf and determine whether $\mathbf{x}^q$ lies within the leaf's corresponding EOA. The outcomes possible from the query are typically:

\begin{enumerate}
    \item \textit{Primary Retrieve}: If the query point lies within the leaf's corresponding EOA, the solution to $\mathbf{f}(\mathbf{x}^q)$ is calculated via linear extrapolation from the solution $\mathbf{f}(\mathbf{x}^0)$ stored at the leaf's center:
    \begin{equation}
        \mathbf{f}(\mathbf{x}^q) \approx \mathbf{f}(\mathbf{x}^0) + \mathbf{J}^0(\mathbf{x}^q - \mathbf{x}^0)
    \end{equation}
    \noindent where $\mathbf{J}^0$ is the Jacobian matrix at the leaf center $\mathbf{x}^0$.

    \item \textit{Grow}: If the query point is not within the leaf's EOA, then an explicit evaluation of $\mathbf{f}(\mathbf{x}^q)$ is done. The difference between $\mathbf{f}(\mathbf{x}^q)$ and $\mathbf{f}(\mathbf{x}^0)$ is then compared, and if it is lower than the user-specified tolerance, the EOA is grown to cover the query point $\mathbf{x}^q$.

    \item \textit{Add}: If the difference between $\mathbf{f}(\mathbf{x}^q)$ and $\mathbf{f}(\mathbf{x}^0)$ is not below the user-specified tolerance, then a new leaf is added at $\mathbf{x}^q$, and its Jacobian $\mathbf{J}^0$ is computed.
\end{enumerate}
Additional details on ISAT can be found in Refs.~\cite{pope1997computationally,lu2009improved}.

\begin{figure}
\centering
\resizebox*{14cm}{!}{\includegraphics{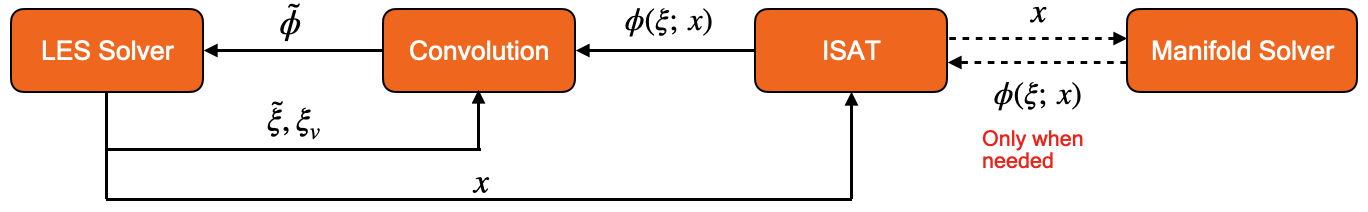}}
\caption{ISAM algorithm coupled to LES.} \label{fig:isam}
\end{figure}

\subsection{In-Situ Adaptive Manifolds (ISAM)} \label{ssec:isam}

Figure \ref{fig:isam} demonstrates how the calculation of the manifold solutions is coupled to LES using ISAM. At the core of ISAM is ISAT \cite{pope1997computationally,lu2009improved}, which is used to store manifold solutions and their Jacobians at the leaves of a binary tree as described in Sec.~\ref{ssec:isat}. In Fig.~\ref{fig:isam}, $\phi(\xi;\,x)$ represents the thermochemical state as a function of the manifold coordinates $\xi$ and the manifold parameters $x$. Numerically, the thermochemical state is discretized with respect to the manifold coordiantes $\xi$. At each LES timestep, the manifold parameters $x$ are calculated for each LES grid point, and ISAT is queried. If ISAT cannot execute a primary retrieve, the manifold solver is called explicitly and the database is updated, as explained in Sec.~\ref{ssec:isat}. ISAT then returns the thermochemical state as a function of the manifold coordinates, and this is passed to convolution-on-the-fly (COTF) \cite{perry2019effect}, which convolves the unfiltered thermochemical state against a presumed subfilter PDF to construct the filtered thermochemical state:

\begin{equation}
    \widetilde{\phi} = \int \phi(\xi;\,x) \widetilde{P}(\xi;\,\widetilde{\xi}, \xi_v) \,d\xi.
\end{equation}

\noindent $\widetilde{P}$ is the subfilter PDF, $\widetilde{\xi}$ is the vector of filtered manifold coordinates, and $\xi_v$ is the manifold coordinates subfilter covariance matrix. The filtered thermochemical state is then used to continue the LES.

ISAM has been previously implemented and evaluated with a variety of canonical flames \cite{lacey2021situ,novoselov2021large,yao2025situ}. Overall, ISAM has been shown that it can achieve similar steady-state timer per timestep compared to traditional pretabulation approaches \cite{lacey2021situ,novoselov2021large,yao2025situ}, but, with larger mechanisms or higher-dimensional manifolds, there is a larger cost associated with the first few timesteps of the simulations as the databases are being populated \cite{novoselov2021large,yao2025situ}.

\subsection{Neural-ISAM} \label{ssec:nisam}

As model complexity grows, the ISAT database memory requirements will eventually become excessive. Therefore, the original ISAM approach will have some limit on model complexity that could be achieved. To address this limitation, in Neural-ISAM, the binary trees are periodically pruned to reduce the memory requirements and accommodate more complex models. At user-specified intervals, the binary tree is searched to identify candidate nodes for pruning (discussed in detail in Section \ref{sssec:id}), neural networks are trained on candidate regions (discussed in detail in Section \ref{sssec:train}), and the binary tree is pruned at each candidate node.

\subsubsection{Identifying pruning candidates} \label{sssec:id}

Candidate regions for pruning should be regions of the database that have high coverage of the input space, meaning there is a large amount of data present in the ISAT database for that region. This ensures that the training data for the neural networks is comprehensive enough such that the neural network will accurately represent the pruned portion of the binary tree. To quantify the coverage of input space for a region of an ISAT database, the coverage density $\rho$ is defined as

\begin{equation}
    \rho = R_n \frac{\sum_j V_{\mathrm{EOA},j}}{V_{\mathrm{AABB}}},
    \label{eq:rho}
\end{equation}

\noindent where $\sum_j V_{\mathrm{EOA},j}$ is the sum of the volumes of the EOAs beneath a node of the binary tree, $V_{\mathrm{AABB}}$ is the volume of the Axis-Aligned Bounding Box (AABB) that covers all EOAs below the node, and $R_n$ is a dimensionality correction term, where $n$ is the dimension of the manifold space. The coverage density is larger for regions of the database that have a higher density of EOAs and is smaller for regions that are sparser.

Each EOA $E$ is defined as

\begin{equation}
    E \equiv  \{ x \,|\, ||L^T(\mathbf{x} - \mathbf{c})|| \le 1\},
\end{equation}

\noindent where $L$ is the lower triangular matrix from the Cholesky decomposition of the Jacobian matrix, and $\mathbf{c}$ is the center of the ellipsoid \cite{pope1997computationally}. With this definition of the EOAs, the volume of the $j$th EOA is found by a dilatation of the volume of a unit sphere with dimension $n$ by the length of the principal axes:

\begin{equation}
    V_{\mathrm{EOA},j} = \frac{\pi^{n/2}}{\Gamma(n/2 +1)} \cdot \frac{1}{| L(j) |}
\end{equation}

\noindent where $| L(j) |$ is the determinant of $L$ for the $j$th EOA and $\Gamma$ is the gamma function. Then, each $V_{\mathrm{EOA},j}$ can be summed to find the total volume occupied by a set of EOAs below a node of the binary tree. 

The AABB that covers all EOAs is calculated by finding the bounding box of each EOA, and taking the minimum and maximum over each local AABB extrema in all axis directions. The AABB coordinates for the $j$th EOA are found as

\begin{equation}
    (x_{i,\min},\,\,x_{i,\max})_j = \left(c_i - \sqrt{L(j)_i^TL(j)_i}\,\,,\,\,c_i + \sqrt{L(j)_i^TL(j)_i}\right),
\end{equation}

\noindent where $(x_{i,\min},\,\,x_{i,\max})_j$ are the bounding box extrema for each dimension $i=1,n$ for the $j$th EOA and $L(j)_i$ denotes the $i$th row of matrix $L$ for the $j$th EOA. After the extrema of the AABB covering all EOAs are found, the volume of the AABB is calculated.

Finally, as dimensionality increases, an ellipsoid occupies less volume relative to the hyperrectangle bounding it. A dimensionality correction term $R_n$ is included in Fig.~\ref{eq:rho} to normalize the volume ratio as the manifold dimension increases and is the ratio of the volume of the unit cube to the unit sphere in dimension $n$:

\begin{equation}
    R_n = \frac{2^n\Gamma(n/2+1)}{\pi^{n/2}}.
\end{equation}

\noindent This ensures the consistency of the definition of the coverage density with varying manifold dimension.

The user specifies a coverage density threshold $\rho_{\mathrm{thresh}}$, which is the minimum coverage density a node of the binary tree can have in order to be considered a candidate node for pruning. An important note is the coverage density of lower nodes in the tree tends to be high due to the small number of EOAs below the node, and pruning these lower nodes does not remove significant portions of the binary tree. Therefore, an additional threshold is included as the minimum number of leaves a node should have in order to be considered a candidate ($N_{L,\mathrm{thresh}}$). If both thresholds are satisfied at a given node ($\rho > \rho_{\mathrm{thresh}}$ and $N_L > N_{L,\mathrm{thresh}}$), the node is pruned. This removes the data structures within ISAT beneath the node, and a trained neural network is assigned to the node in their place. The process of training these neural networks for nodes selected for pruning is described in the following section.

\subsubsection{Training the neural networks} \label{sssec:train}

\begin{figure}
\centering
\resizebox*{6cm}{!}{\includegraphics{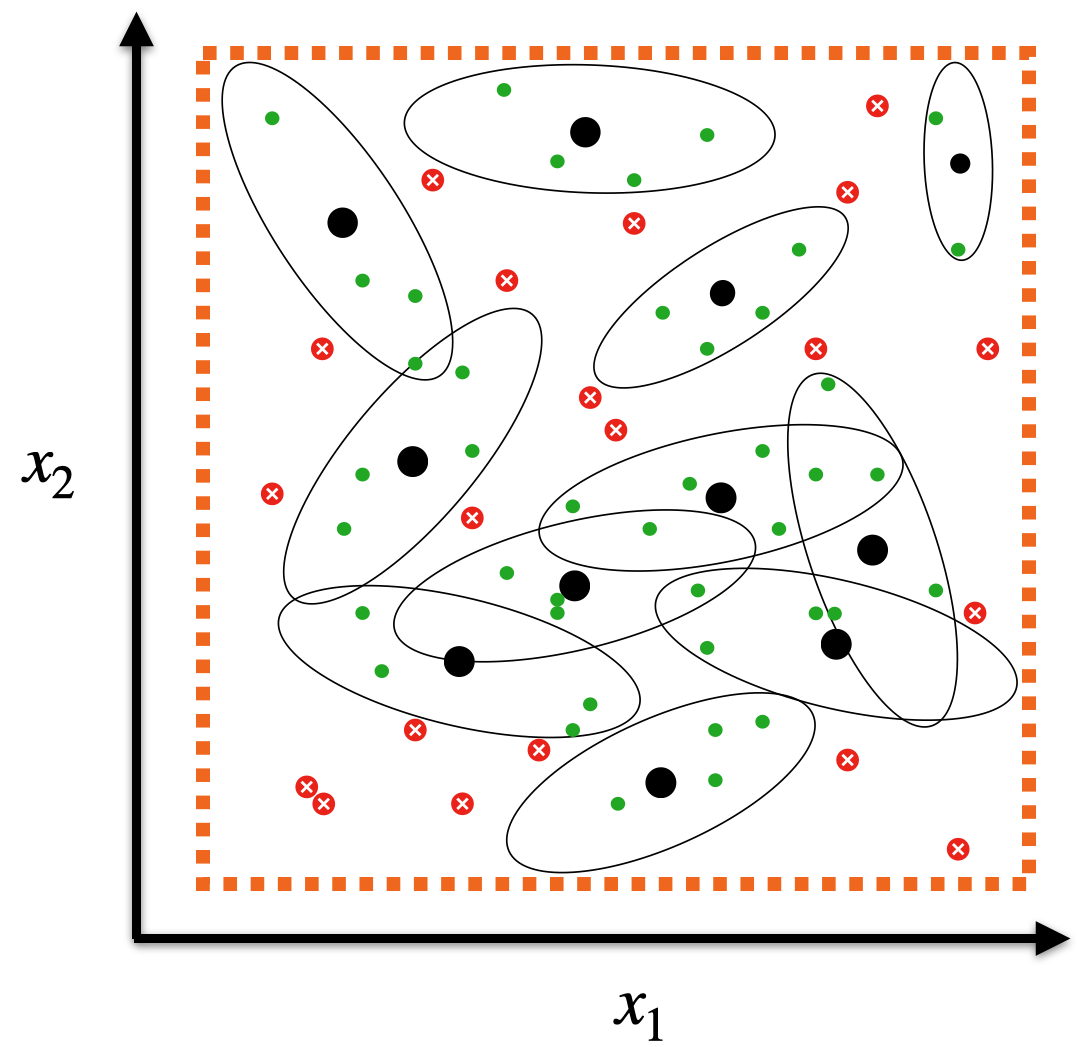}}
\caption{Diagram demonstrating the training data generation process for Neural-ISAM. Black dots represent the leaf centers, black ellipses are the EOAs, and the orange dotted box is the candidate prune region (AABB). The green dots within the EOAs are points accepted for training, while the red X's outside of the EOAs are points rejected from training.} \label{fig:nisam}
\end{figure}

Once the binary tree has been searched and the candidate nodes have been flagged for pruning, the training data for the neural networks that will replace the candidate nodes is generated by sampling random input values within the bounding box of each candidate node:
\begin{enumerate}
    \item If a sampled point lies within an EOA of a leaf below the candidate node, the sample point is kept.
    \item If a sampled point does not lie within an EOA of a leaf below the candidate node, the sample point is rejected.
\end{enumerate}
This sampling continues until the number of accepted points reaches a user-specified number of training samples. Since the gathered points all lie within an EOA of a leaf in the candidate prune region, the thermochemical states are calculated via linear extrapolation from their corresponding leaf, requiring no additional explicit manifold calculations to generate the training dataset. Figure \ref{fig:nisam} demonstrates this process for a sample two-dimensional input space. After the training data has been generated, the training tasks for the candidate nodes across every MPI task are distributed evenly among all of the MPI processes to ensure efficient training. This means that multiple training tasks can occur simultaneously. 

\begin{figure}
\centering
\resizebox*{\linewidth}{!}{\includegraphics{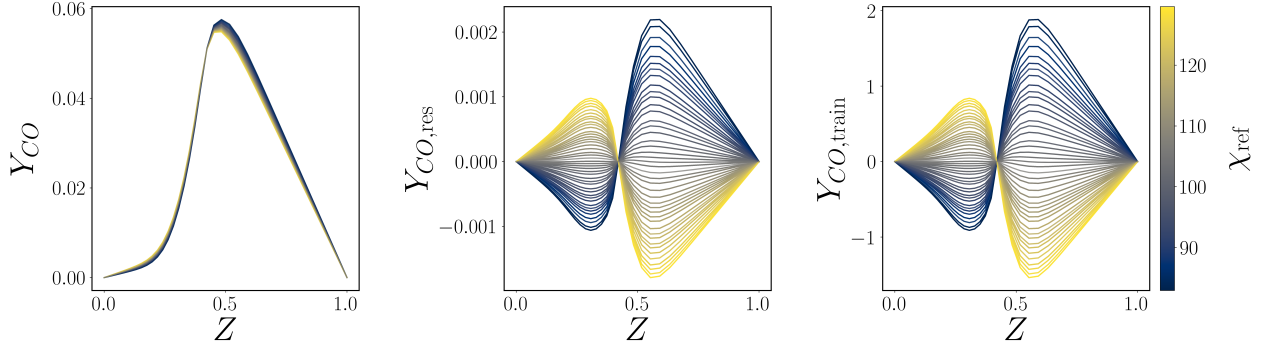}}
\caption{Sample training data profiles before and after the scaling transformation, shown for $Y_{CO}$ for Sandia Flame D.} \label{fig:tdata}
\end{figure}

The neural networks are tasked with predicting $\phi(\xi)$, which is thermochemical state discretized over $N_{\xi}$ total points in the manifold coordinates. With $N_{\mathrm{var}}$ thermochemical quantities being predicted, this results in $N_{\xi} \cdot N_{\mathrm{var}}$ output values that need to be predicted by the neural network (stacked into a single array). In order to train the neural networks, the data needs to be transformed due to the large range of scales present in the thermochemical state. The data is transformed as follows, and an example visualization is shown in Figure \ref{fig:tdata}:

\begin{enumerate}
    \item Mean profiles $\bar{\phi}(\xi)$ over all training samples are found for each quantity at each $\xi$ point, which is then subtracted from the training profiles to get the residual from the mean $\phi_{\mathrm{res}}(\xi)$:
    \begin{equation}
        \phi_{\mathrm{res}}(\xi) = \phi(\xi) - \bar{\phi}(\xi)\,.
    \end{equation}
    \item The residual profiles are then scaled by the maximum standard deviation for each thermochemical quantity across the training data. More specifically, the standard deviation of the training data is calculated for each $\xi$ point across all training samples, and, for each thermochemical quantity, the maximum standard deviation $\sigma_\phi$ over all $\xi$ points is used as the scaling value:
    \begin{equation}
        \phi_{\mathrm{train}}(\xi) = \phi_{\mathrm{res}}(\xi) / \sigma_{\phi}\,.
    \end{equation}
\end{enumerate}

\noindent The mean profiles $\bar{\phi}(\xi)$ and the standard deviation scale values $\sigma_{\phi}$ used to transform the training data are stored with the neural network to rescale the predictions at inference time.

After the data is scaled, each training task uses Bayesian Optimization \cite{omalley2019kerastuner} to tune the number of layers and the number of neurons used in each layer. Since choosing the number of neurons in each layer constitutes a very large parameter space for the Bayesian optimizer, the approach that empirically works best with the Neural-ISAM implementation is to instead optimize an expansion factor $\alpha$ that encodes the number of neurons in layer $i$:

\begin{equation}
    N_{\mathrm{neurons},i} = \min(\alpha^i, N_{\mathrm{out}})
\end{equation}

\noindent where $N_{\mathrm{neurons},i}$ is the number of neurons in layer $i$ and $N_{\mathrm{out}}$ is the number of output neurons. The range of values explored with Bayesian Optimization for the neural networks was $2-8$ for the number of layers and $1.5-5.0$ for the expansion factor $\alpha$. For each training task, the Bayesian optimizer was run for 25 trials, which was found sufficient to converge to an optimal architecture for the current implementation. During the training, a validation dataset 20\% the size of the training dataset is used. A learning rate of $7\times10^{-3}$ was used, and each training operation had a maximum number of epochs of 500 with early stopping if the validation loss did not continue to decrease after 20 epochs. The Adam optimizer was used, and leaky ReLU was empirically found to be the best activation function for these applications.

\section{Test Cases} \label{ssec:test}

For each simulation detailed in the following subsections, a full ISAT database was used from the start, meaning there were very few adds or grows occurring during each timestep. The ISAT database was built with a relative error tolerance of $1\times10^{-4}$ and an absolute error tolerance of $1\times10^{-10}$. The purpose of this is to compare the cost and performance of Neural-ISAM relative to the steady-state behavior of ISAM. Neural-ISAM databases were traversed every 50 timesteps to identify any candidate nodes for pruning. The pruning for all candidate nodes are executed when they are identified. The following subsections detail the manifolds used for each test case, as well as the computational setup of each simulation.

\begin{table}
\centering
\caption{Simulation cases for testing Neural-ISAM. The number after ``r" corresponds to the coverage density threshold $\rho_{\mathrm{thresh}}$, and the number after ``N" corresponds to the number of leaves threshold $N_{L,\mathrm{thresh}}$.}{
\begin{tabular}{lccccc}
\toprule
\multicolumn{3}{c}{Sandia Flame D} & \multicolumn{3}{c}{Sandia Sooting Flame} \\
\cmidrule(lr){1-3}\cmidrule(lr){4-6}
Name & $\rho_{\mathrm{thresh}}$ & $N_{L,\mathrm{thresh}}$
    & Name & $\rho_{\mathrm{thresh}}$ & $N_{L,\mathrm{thresh}}$ \\ 
\midrule
noprune & N/A & N/A & noprune & N/A & N/A \\
r0.2N10 & 0.2 & 10 & r0.6N90  & 0.6 & 90 \\
r0.2N40 & 0.2 & 40 & r0.6N350 & 0.6 & 350 \\
r0.6N10 & 0.6 & 10 & r0.8N90  & 0.8 & 90 \\
r0.6N40 & 0.6 & 40 & r0.8N350 & 0.8 & 350 \\
\bottomrule
\end{tabular}}
\label{tab:cases}
\end{table}

\subsection{Sandia Flame D}

LES of Sandia Flame D \cite{barlow1998effects} is used as a proof of concept to demonstrate the performance of the Neural-ISAM method. A nonpremixed manifold is used identical to the formulation of Ref. \cite{lacey2021situ}. For the nonpremixed manifold, the manifold coordinate is the mixture fraction $Z$, and the manifold parameter is the mixture fraction dissipation rate $\chi$. The manifold equations are defined \cite{peters1984laminar}:
\begin{equation}
    0 = \frac{\rho \chi}{2} \frac{d^2Y_k}{dZ^2} + \dot{m}_k,
\end{equation}

\noindent where the dissipation rate profile as a function of mixture fraction $Z$ is presumed to be the product of a reference dissipation rate $\chi_{\mathrm{ref}}$ and a normalized profile \cite{peters1983local}:
\begin{equation}
    \chi = \chi_{\mathrm{ref}} \cdot \frac{f(Z)}{f(Z_{\mathrm{ref}})},
\end{equation}

\noindent and the stoichiometric mixture fraction is used for $Z_{\mathrm{ref}}$. The thermochemical state $\phi(Z)$ is now defined as a function of the mixture fraction $Z$, and the manifold solutions are parameterized by the reference mixture fraction dissipation rate $\chi_{\mathrm{ref}}$ \cite{lacey2021situ,peters1983local}.

The simulation setup is as identical to Ref. \cite{lacey2021situ} with a $256\times144\times64$ stretched grid in cylindrical coordinates. NGA, a structured, finite-difference, low Mach number flow solver \cite{desjardins2008high,macart2016semi} was used for the LES, and PDRs \cite{mueller2020physically, pdrs} was used for solving the manifold equations with the GRI-3.0 mechanism \cite{gri3.0}. Table \ref{tab:cases} lists the range of values of $\rho_{\mathrm{thresh}}$ and $N_{L,\mathrm{thresh}}$ that were considered. The selected values of $\rho_{\mathrm{thresh}}$ were $0.2$ and $0.6$ because there were no nodes of the tree satisfying the pruning criteria below $\rho_{\mathrm{thresh}}=0.2$ and above $\rho_{\mathrm{thresh}}=0.6$. The values of $N_{L,\mathrm{thresh}}$ were selected as 10 and 40 because each of these values corresponds to 5\% and 20\% of the total size of the unpruned binary tree respectively, which were found to best demonstrate the behavior of Neural-ISAM.

The manifold for this flame is parameterized only by the reference mixture fraction dissipation rate $\chi_{\mathrm{ref}}$, resulting in a one dimensional input space for the ISAT database. The number of output thermochemical variables to be predicted was $N_{\mathrm{var}}=11$ (density, diffusivity, viscosity, temperature, and seven mass fractions) and the number of profile points used in mixture fraction was $N_Z=33$, resulting in the neural networks predicting $N_Z \cdot N_{\mathrm{var}} = 363$ outputs for each input value.

\subsection{Sandia Sooting Flame}

To further test the Neural-ISAM method on a more complex manifold, the Sandia sooting ethylene/air piloted turbulent jet flame \cite{zhang2011design} was run as detailed in Ref. \cite{yao2025situ}. The manifold model is extended from the previous section to account for radiation heat losses and soot. The same manifold equations are considered but with additional terms for radiation heat losses \cite{nunno2019manifold} and soot \cite{mueller2012model,yao2025situ}. The manifold model is parameterized by the same mixture fraction reference dissipation rate $\chi_{\mathrm{ref}}$ as well as an additional radiation factor $\Omega$ \cite{nunno2019manifold}.

Simulations were run with a $192\times92\times36$ stretched grid in cylindrical coordinates. Similarly, NGA \cite{desjardins2008high,macart2016semi} is used for the LES, and PDRs \cite{mueller2020physically, pdrs} is used for the manifold calculations with the reduced mechanism from Ref. \cite{bisetti2012formation} that considers PAH chemistry up to naphthalene. Table \ref{tab:cases} lists the cases run for the Sandia sooting flame. The selected values of $\rho_{\mathrm{thresh}}$ were $0.6$ and $0.8$ because $\rho_{\mathrm{thresh}}<0.6$ resulted in poor neural network predictions that inhibited the evolution of the simulation, and nodes with $\rho_{\mathrm{thresh}}>0.8$ were rare in the ISAT databases. The values of $N_{L,\mathrm{thresh}}$ were selected as 90 and 350 because each of these values corresponds to approximately 5\% and 20\% of the total size of the unpruned binary tree, similar to the Sandia Flame D simulations.

The additional manifold parameter $\Omega$ increases the ISAT database dimension by one. The number of output thermochemical variables also is increased for this scenario to $N_{\mathrm{var}}=34$ to also include the heat loss parameter $H$ (essentially an enthalpy deficit~\cite{yao2025situ}) and variables important for soot formation and evolution, and the number of $Z$ points also is increased to $N_Z=65$. This results in the neural networks predicting $N_Z \cdot N_{\mathrm{var}} = 2210$ outputs for each input pair. 

Additional data transformation has to be done for this test case, for the heat loss parameter is highly sensitive to the inputs and occupies a large range of scales. The standard data transformation described in Section \ref{sssec:train} alone does not transform this quantity into an easy learnable profile spread as previously shown in Figure \ref{fig:tdata}, and using log scaling diminishes small variations in $H$ that are important to the manifold model's performance. Instead, the profiles for $H$ are scaled as
\begin{equation}
    H' = \mathrm{arcsinh}(H/c),
\end{equation}
\noindent where $c$ is calculated as the median value of $H$ across the entire training dataset. Scaling with arcsinh has been shown to be effective for quantities that span large ranges of scales, with the small scale behavior still being important \cite{burbidge1988alternative}. This function is also easily inverted at inference time to unscale predictions and recover the actual profile for $H$. Additionally, because the underlying algorithm of incorporating non-adiabatic effects into the manifold is heavily dependent on the prediction of $H$ \cite{yao2025situ}, the loss function weights were increased to 2.0 for all values of $Z$ for the quantity $H$, with the loss weights remaining at 1.0 for all other quantities.

\section{Results}

\subsection{Sandia Flame D}

Memory usage for the Sandia Flame D simulations is shown in the left half of Table~\ref{tab:sd-mem}. There is a reduction in memory usage of the databases for any level of pruning. For higher $N_{L,\mathrm{thresh}}$, a more significant reduction in memory around 20\% is achieved as more leaves are being pruned from the binary tree. Therefore, databases pruned with a lower $N_{L,\mathrm{thresh}}$ resulted in a memory reduction of only approximately 14\%. The explanation is simple: with a threshold for fewer leaves, pruning occurs deeper in the tree and removes fewer leaves. There is not much difference in memory usage between high $\rho_{\mathrm{thresh}}$ and low $\rho_{\mathrm{thresh}}$ for the Sandia D simulations, for the one-dimensional manifold input space means the coverage density threshold is not as restrictive as in higher-dimensional input spaces.

\begin{figure}[t!]
\centering
\resizebox*{\linewidth}{!}{\includegraphics{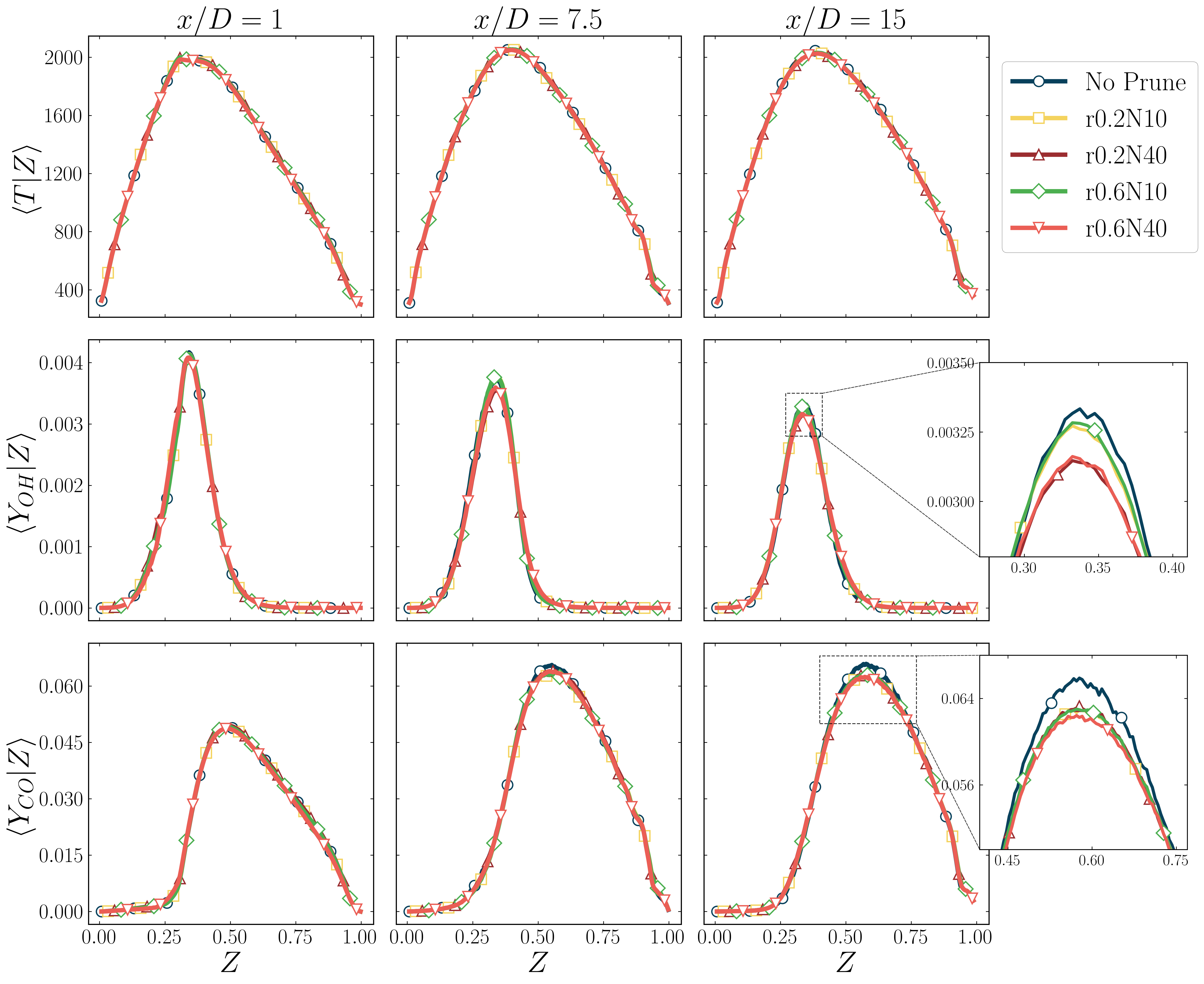}}\\
\caption{Conditional statistics from LES for each Sandia Flame D test case shown for temperature (top), OH mass fraction (middle),  and CO mass fraction (bottom).} \label{fig:sdst}
\end{figure}

Figure~\ref{fig:sdst} shows the conditional statistics for temperature and CO mass fraction for each simulation case. Between all simulation cases the statistics are nearly identical, demonstrating the accuracy of the pruned databases is maintained in the flame statistics. Only slight deviations are observed in the predictions of $Y_{OH}$ and $Y_{CO}$. For $Y_{OH}$, all Neural-ISAM simulations except r0.6N10 slightly under predict the conditional mean for various distances downstream from the burner. For $Y_{CO}$, all Neural-ISAM simulations slightly under predict the CO mass fraction far downstream from the burner. The deviations are quite small and the overall accuracy of the simulations is very high. That said, Neural-ISAM with more aggressive pruning does tend to be slightly less accurate compared to less aggressive pruning.

\begin{figure}[t]
\centering
\resizebox*{14cm}{!}{\includegraphics{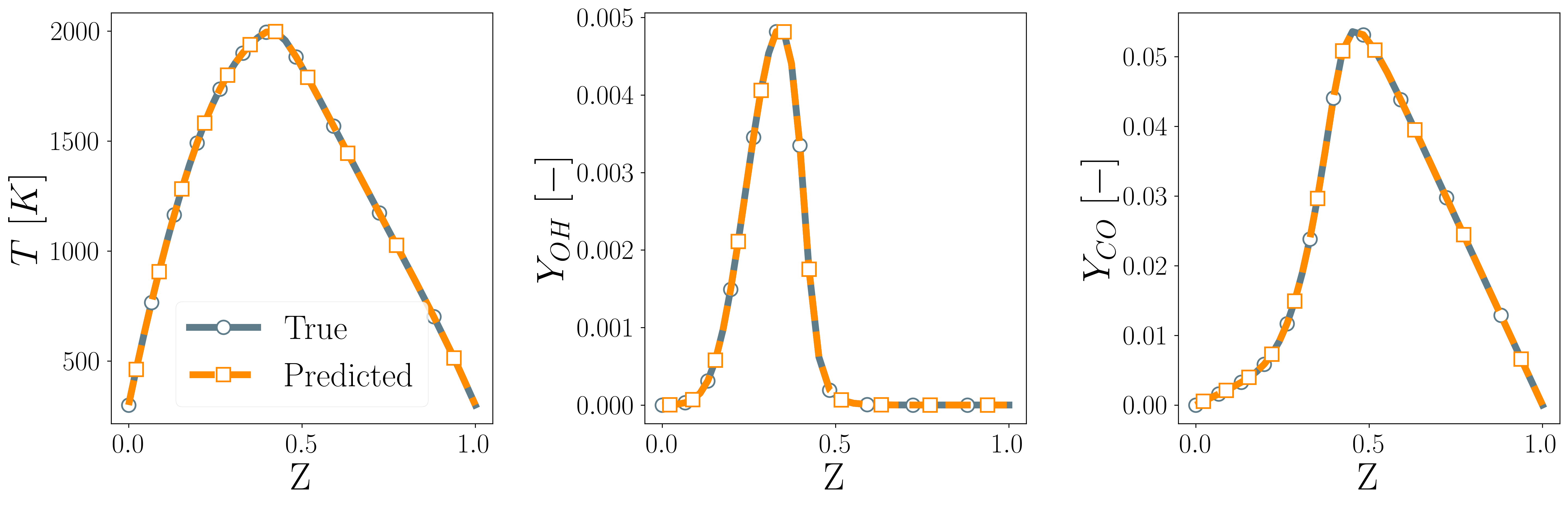}}\\
\caption{Sample NN predictions from the ISAT database generated in the Sandia D r0.6N40 simulation compared to the ground truth manifold solution for $\chi_{\mathrm{ref}} = 168.9{\:\rm s}^{-1}$.} \label{fig:preds}
\end{figure}

To further demonstrate the accuracy of predictions, sample neural network predictions from the ISAT database for case r0.6N40 are compared to the true manifold solution in Fig.~\ref{fig:preds}. It is clearly shown that the predictions of the neural networks align well with the true manifold solutions for this problem. While predictions and statistics are only shown here for temperature, OH mass fraction, and CO mass fractions, the accuracy of the neural network predictions are consistent across all thermochemical quantities.

\begin{table}[t!]
\centering
\caption{ISAT database memory comparison for each test flame (across all MPI processes).}{
\begin{tabular}{cccc}
\toprule
\multicolumn{2}{c}{Sandia Flame D} & \multicolumn{2}{c}{Sandia Sooting Flame} \\
\cmidrule(lr){1-2}\cmidrule(lr){3-4}
Name & ISAT database memory
    & Name & ISAT database memory \\ 
\midrule
noprune & 988 MB & noprune  & 9.3 GB\\
r0.2N10 & 861 MB & r0.6N90  & 12.0 GB\\
r0.2N40 & 792 MB & r0.6N350 & 5.7 GB\\
r0.6N10 & 834 MB & r0.8N90  & 11.0 GB\\
r0.6N40 & 737 MB & r0.8N350 & 6.1 GB\\
\bottomrule
\end{tabular}}
\label{tab:sd-mem}
\end{table}

\begin{table}[t!]
\centering
\caption{Average time per timestep for each Sandia Flame D simulation case in seconds.}{
\begin{tabular}{cccc}
\toprule
Name & Time/timestep (before pruning) & Training time & Time/timestep (after pruning) \\
\hline
noprune & 4.94 & N/A      & N/A \\
r0.2N10 & 5.39 & 5911.71 & 5.22 \\
r0.2N40 & 5.22 & 2086.12 & 8.98 \\
r0.6N10 & 5.08 & 6633.05 & 5.50 \\
r0.6N40 & 5.21 & 2107.32 & 8.80 \\
\hline
\end{tabular}}

\label{tab:sd-time}
\end{table}

\begin{table}[t!]
\centering
\caption{Evaluation statistics for NNs in each run case for both Sandia Flame D and the Sandia Sooting flame. ``NN evals" is the average number of NN evaluations done per timestep after training.}{
\begin{tabular}{cccccc}
\toprule
\multicolumn{3}{c}{Sandia Flame D} & \multicolumn{3}{c}{Sandia Sooting Flame} \\
\cmidrule(lr){1-3}\cmidrule(lr){4-6}
Name & Number of NNs & NN evals.
    & Name & Number of NNs & NN evals. \\ 
\midrule
r0.2N10 & 545 & 223,034 & r0.6N90  & 1089 & 543,446 \\
r0.2N40 & 146 & 353,152 & r0.6N350 & 372  & 772,710 \\
r0.6N10 & 545 & 209,809 & r0.8N90  & 1077 & 522,260 \\
r0.6N40 & 146 & 349,871 & r0.8N350 & 372  & 752,801 \\
\bottomrule
\end{tabular}}
\label{tab:nnstat}
\end{table}

Finally, Table \ref{tab:sd-time} shows computational timing for each simulation case including time per timestep before and after pruning, and training time for the Neural-ISAM simulations. Comparing simulations with $N_{L,\mathrm{thresh}}=10$ to simulations with $N_{L,\mathrm{thresh}}=40$, it can be seen that the training time is longer. For the $N_{L,\mathrm{thresh}}=10$ simulations, there are a larger number of NNs to train, which is shown explicitly in the left half of Table \ref{tab:nnstat}. For time per timestep after training, it can be seen that simulations with lower $N_{L,\mathrm{thresh}}$ have comparable timing to the baseline time per timestep, while simulations with higher $N_{L,\mathrm{thresh}}$ have increased time per timestep. This is due to the fact that higher $N_{L,\mathrm{thresh}}$ simulations have an increased number of NN evaluations, as shown in the left half of Table \ref{tab:nnstat}. After the ISAT database is pruned, what would be primary retrieves from a standard ISAT database become neural network evaluations. For this specific problem, a single primary retrieve takes approximately $5{\rm \mu s}$ while a single neural network evaluation takes approximately $10{\rm \mu s}$. The neural network evaluations take slightly longer as there are more matrix multiplications than the linear extrapolation used in a primary retrieve.

\subsection{Sandia Sooting flame}

For the Sandia Sooting flame, the memory of the final ISAT databases are shown in the right half of Table \ref{tab:sd-mem}. For the simulations with high $N_{L,\mathrm{thresh}}=350$, the database memory has been significantly reduced compared to the baseline case without any pruning (a reduction by approximately 35\%). It is also important to note that this reduction in memory compared to the baseline is more significant than what was seen for Sandia Flame D. However, for the simulations with lower $N_{L,\mathrm{thresh}}=90$, the final memory of the ISAT databases are larger than the baseline case. This is due to the much larger number of NNs occupying the database after pruning with a low $N_{L,\mathrm{thresh}}$, and the necessarily larger model size of the NNs to accommodate the increased number of thermochemical variables for the more complex manifold model. Therefore, for a lower $N_{L,\mathrm{thresh}}$, the memory usage of the database with NNs can actually be larger than the original model if $N_{L,\mathrm{thresh}}$ is set too low. 

\begin{figure}[t!]
\centering
\resizebox*{14cm}{!}{\includegraphics{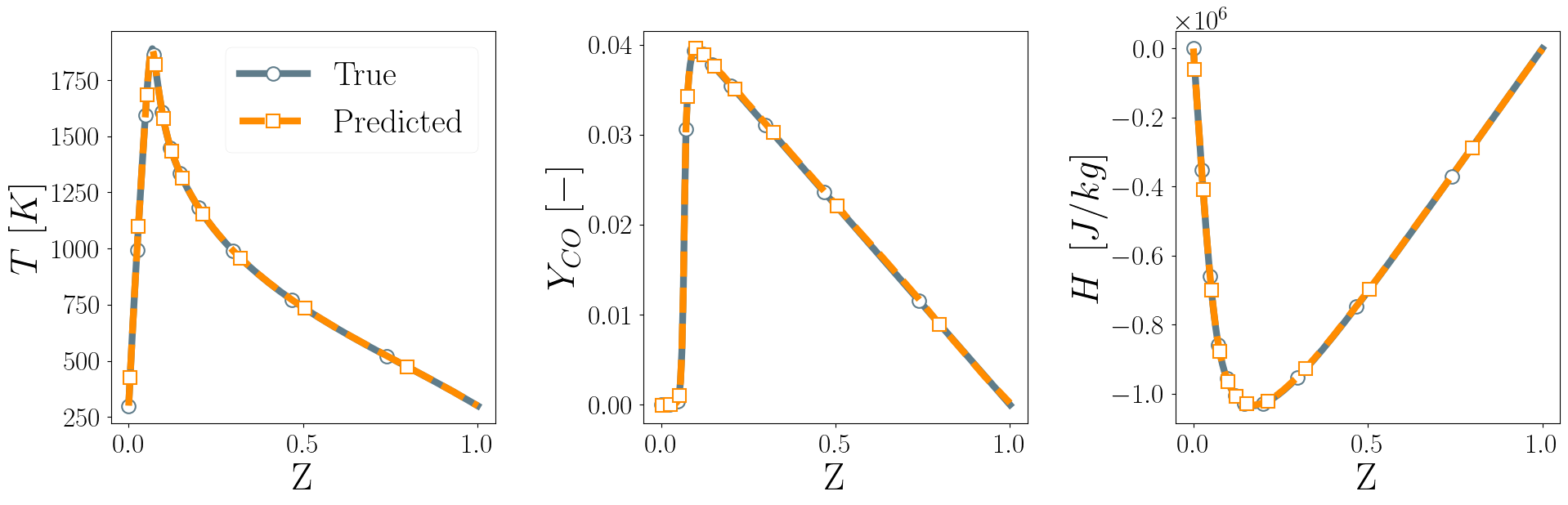}}\\
\resizebox*{14cm}{!}{\includegraphics{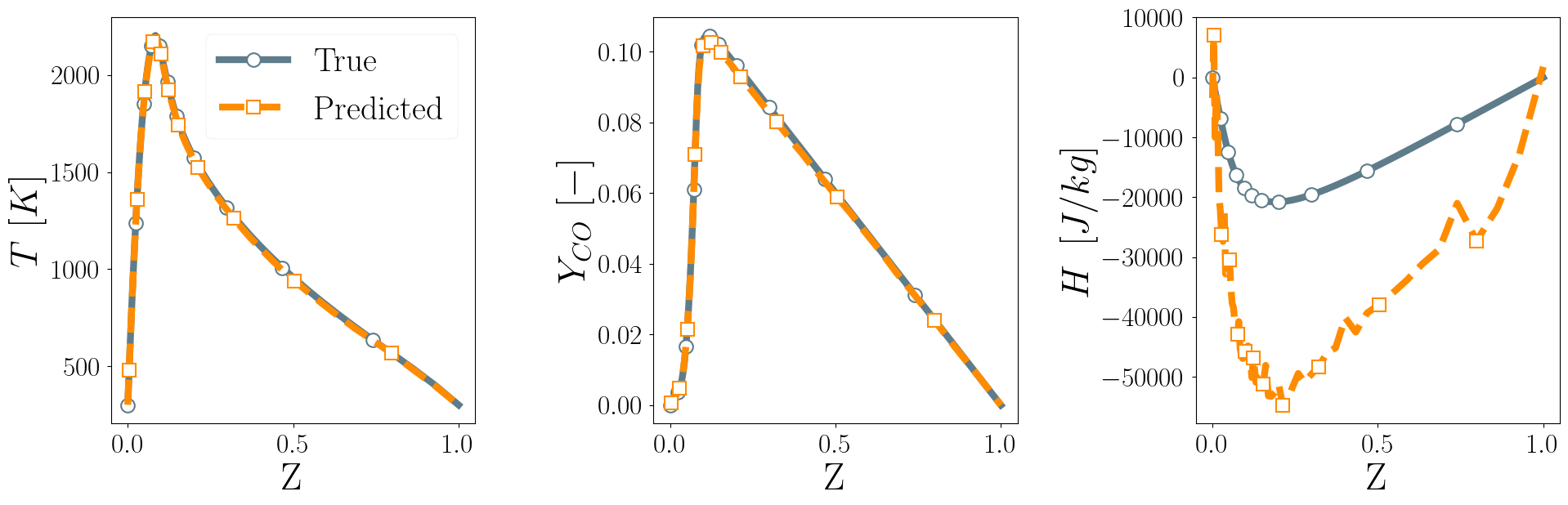}}\\
\caption{Sample NN predictions from the Sandia Sooting flame. The top row is taken from the r0.8N90 simulation for $\chi_{\mathrm{ref}}=0.04\:{\rm s}^{-1}$ and $\Omega=0.309$, while the bottom row is taken from the r0.6N350 simulation for $\chi_{\mathrm{ref}}=5.02\:{\rm s}^{-1}$ and $\Omega=0.505$. Temperature, $CO$ mass fraction, and heat loss parameter are plotted.} \label{fig:soot-preds}
\end{figure}

Figure \ref{fig:soot-preds} shows predictions from the pruned databases for r0.8N90 (a database with highly accurate solutions after training) in the top row and r0.6N350 (a database with less accurate solutions after training) in the bottom row. With the more accurate database, the NNs are able to reconstruct the thermochemical profiles very well. As for the less accurate database, when the NNs are asked to predict at points where there was sparse training data, specifically in between the EOAs, the model struggles to predict important thermochemical quantities. The particular example shown is the heat loss parameter $H$, which is central to the operation of the ISAM algorithm with non-adiabatic effects in LES \cite{yao2025situ}. The bottom row of Fig. \ref{fig:soot-preds} demonstrates this for the r0.6N350 simulation as the predicted value shows a significant discrepancy compared to the manifold solution. The lower $N_{L,\mathrm{thresh}}$ and higher $\rho_{\mathrm{thresh}}$ results in better prediction accuracy not only because there is a higher coverage density being enforced, but with a lower $N_{L,\mathrm{thresh}}$ the pruning occurs lower down in the tree. With a pruning region lower in the tree, the neural network has to fit a smaller range of manifold input values, improving its performance. While the predictions are more accurate, as previously discussed, with lower $N_{L,\mathrm{thresh}}$, there is generally lower memory savings. This motivates the development of strategies for decreasing inaccurate predictions in regions of sparse training data, especially in between EOAs, as the manifold complexity and dimension increases.

\begin{table}[t!]
\centering
\caption{Average time per timestep for each Sandia Sooting flame simulation case in seconds.}{
\begin{tabular}{cccc}
\toprule
Name & Time/timestep (before pruning) & Training time & Time/timestep (after pruning) \\
\hline
noprune  & 12.50 & N/A      & N/A \\
r0.6N90  & 13.14 & 14953.30 & 57.75 \\
r0.6N350 & 12.36 & 5081.36  & 116.19 \\
r0.8N90  & 13.89 & 14818.10 & 29.87 \\
r0.8N350 & 14.24 & 4976.19  & 84.71 \\
\hline
\end{tabular}}

\label{tab:soot-time}
\end{table}

Table \ref{tab:soot-time} shows computational timing for the Sandia Sooting Flame simulations. Similar to the Sandia Flame D simulations, for the low $N_{L,\mathrm{thresh}}=90$ simulations the training time is significantly higher compared to the high $N_{L,\mathrm{thresh}}=350$ case due to the large number of NNs being trained, as shown in the right half of Table \ref{tab:nnstat}. Additionally, the time per timestep after pruning for the simulations with $N_{L,\mathrm{thresh}}=350$ are much higher than for the $N_{L,\mathrm{thresh}}=90$ cases. Once again this is due to the higher number of NN evaluations being done each timestep, also shown in the right half of Table \ref{tab:nnstat}. This is for the same reason as discussed in the Sandia Flame D results: each neural network evaluation is slightly more computationally expensive than the linear extrapolation of a primary retrieve. Finally, the Sandia Sooting flame demonstrates the difference in performance between cases with low $\rho_{\mathrm{thresh}}$ and high $\rho_{\mathrm{thresh}}$. The time per timestep for the cases with $\rho_{\mathrm{thresh}}=0.8$ are both lower than the corresponding simulation with $\rho_{\mathrm{thresh}}=0.6$. This is due to the coverage density threshold further restricting the number of NNs that are created in the dataset, which in turn further decreases the number of NN evaluations done at each timestep as shown in the right half of Table \ref{tab:nnstat}, ultimately leading to slightly lower timestep durations compared to the lower $\rho_{\mathrm{thresh}}$ value.

\section{Conclusions}

In this work, a novel Neural-ISAM framework was developed to reduce the memory requirements of the original ISAM method as the complexity of manifold-based combustion models increases. The Neural-ISAM method identifies regions of ISAT binary tree databases that can be pruned based on two quantities, the EOA coverage density $\rho$ and the number of leaves below a given node $N_L$, with each having user-specified thresholds for pruning ($\rho_\mathrm{thresh}$ and $N_{L,\mathrm{thresh}}$, respectively). Training data are generated at pruning candidate nodes by sampling the EOAs below the candidate node, the data are transformed and scaled, and a neural network is trained in-situ with the model architecture being selected by Bayesian Optimization. The region of the binary tree below the candidate node is then pruned off of the database, and any further queries in the pruned region are done via neural network evaluations. LES of Sandia Flame D was conducted as a proof of concept, and LES of the Sandia Sooting flame was conducted as a further validation case with a more complex manifold model. Varying values of $\rho_\mathrm{thresh}$ and $N_{L,\mathrm{thresh}}$ were considered for each test flame. 

Results from the Sandia Flame D simulations showed that pruning the ISAT databases and replacing the pruned regions with neural networks decreased the database memory by between 14--20\% depending on the values of $\rho_\mathrm{thresh}$ and $N_{L,\mathrm{thresh}}$. The conditional statistics for each simulation matched the baseline simulation very well with only a slight underprediction of $Y_{CO}$ and $Y_{OH}$ for some values of $\rho_\mathrm{thresh}$ and $N_{L,\mathrm{thresh}}$ that resulted in more aggressive pruning. Analyzing the computational time showed that the main cost is due to the training step, which takes less time when fewer neural networks need to be trained with a larger $N_{L,\mathrm{thresh}}$. The post-training time per timestep also depends on the number of neural network evaluations that are done, with higher numbers of evaluations leading to slightly longer times per timestep.

Results from the Sandia Sooting flame simulations showed that pruning the ISAT databases with a high enough $N_{L,\mathrm{thresh}}$ resulted in an even more significant reduction in memory of approximately 34--38\%. It was also shown that, if $N_{L,\mathrm{thresh}}$ is set too low, the memory benefits of Neural-ISAM are not achieved. The neural network prediction accuracy with the more complicated manifold was best for high $\rho_\mathrm{thresh}$ and low $N_{L,\mathrm{thresh}}$. The lowest accuracy simulations were those with low $\rho_\mathrm{thresh}$ and high $N_{L,\mathrm{thresh}}$. The models struggled to predict the heat loss parameter in regions where the training data was sparse in between EOAs. The computational timing showed a similar trend as Sandia Flame D but further demonstrated that a lower number of neural network evaluations per timestep results in a lower time per timestep.

Overall, between these two test cases of Neural-ISAM, it was shown that pruning regions of ISAT databases and replacing them with neural networks can significantly reduce the memory usage of the manifold databases, especially for more complex manifolds. This comes at an increased computational cost due to the slightly higher evaluation time and large number of neural network evaluations needed. The increased computational cost could be mitigated in the future with more efficient neural network evaluations within ISAT. The accuracy of the predictions were high between the two test cases, with small deviations for some thermochemical quantities in the simple test case and larger prediction errors for sensitive quantities in the more complicated test case depending on the the coverage density and number of leaves thresholds set. This revealed the need for further work to be done to identify where the neural network models perform poorly, especially in regions of sparse training data in between EOAs, and improve their prediction accuracy.

\newpage
\bibliographystyle{unsrt}

\end{document}